\documentclass[10pt,twocolumn,letterpaper]{article}
\usepackage{changes}
\usepackage{cvpr}              
\usepackage[accsupp]{axessibility}
\definecolor{cvprblue}{rgb}{0.21,0.49,0.74}
\usepackage[pagebackref,breaklinks,colorlinks,citecolor=cvprblue,linkcolor=cvprblue,urlcolor=cvprblue]{hyperref}
\usepackage{multirow}

\usepackage{colortbl}%
  \newcommand{\myrowcolour}{\rowcolor[gray]{0.925}}

\def\confName{CVPR}
\def\confYear{2024}

\title{Motion-adaptive Separable Collaborative Filters for Blind Motion Deblurring}

\author{Chengxu Liu\textsuperscript{1,4}\quad\quad Xuan Wang\textsuperscript{2}\quad\quad Xiangyu Xu\textsuperscript{1}\quad\quad Ruhao Tian\textsuperscript{1}\\Shuai Li\textsuperscript{2}\quad\quad Xueming Qian\textsuperscript{1,4}\quad\quad Ming-Hsuan Yang\textsuperscript{3}\\
\textsuperscript{1}Xi’an Jiaotong University \quad \textsuperscript{2}MEGVII Technology\quad
\textsuperscript{3}University of California, Merced \\
\textsuperscript{4}Shaanxi Yulan Jiuzhou Intelligent Optoelectronic Technology Co., Ltd\\
}

\begin{document}

\maketitle

\begin{abstract}
Eliminating image blur produced by various kinds of motion has been a challenging problem. Dominant approaches rely heavily on model capacity to remove blurring by reconstructing residual from blurry observation in feature space. These practices not only prevent the capture of spatially variable motion in the real world but also ignore the tailored handling of various motions in image space. In this paper, we propose a novel real-world deblurring filtering model called the Motion-adaptive Separable Collaborative (MISC) Filter. In particular, we use a motion estimation network to capture motion information from neighborhoods, thereby adaptively estimating spatially-variant motion flow, mask, kernels, weights, and offsets to obtain the MISC Filter. The MISC Filter first aligns the motion-induced blurring patterns to the motion middle along the predicted flow direction, and then collaboratively filters the aligned image through the predicted kernels, weights, and offsets to generate the output. This design can handle more generalized and complex motion in a spatially differentiated manner. Furthermore, we analyze the relationships between the motion estimation network and the residual reconstruction network. Extensive experiments on four widely used benchmarks demonstrate that our method provides an effective solution for real-world motion blur removal and achieves state-of-the-art performance. Code is available at \url{https://github.com/ChengxuLiu/MISCFilter}.

\end{abstract}

\begin{figure}[t]
  \centering
    \includegraphics[width=1.0\linewidth,page=1]{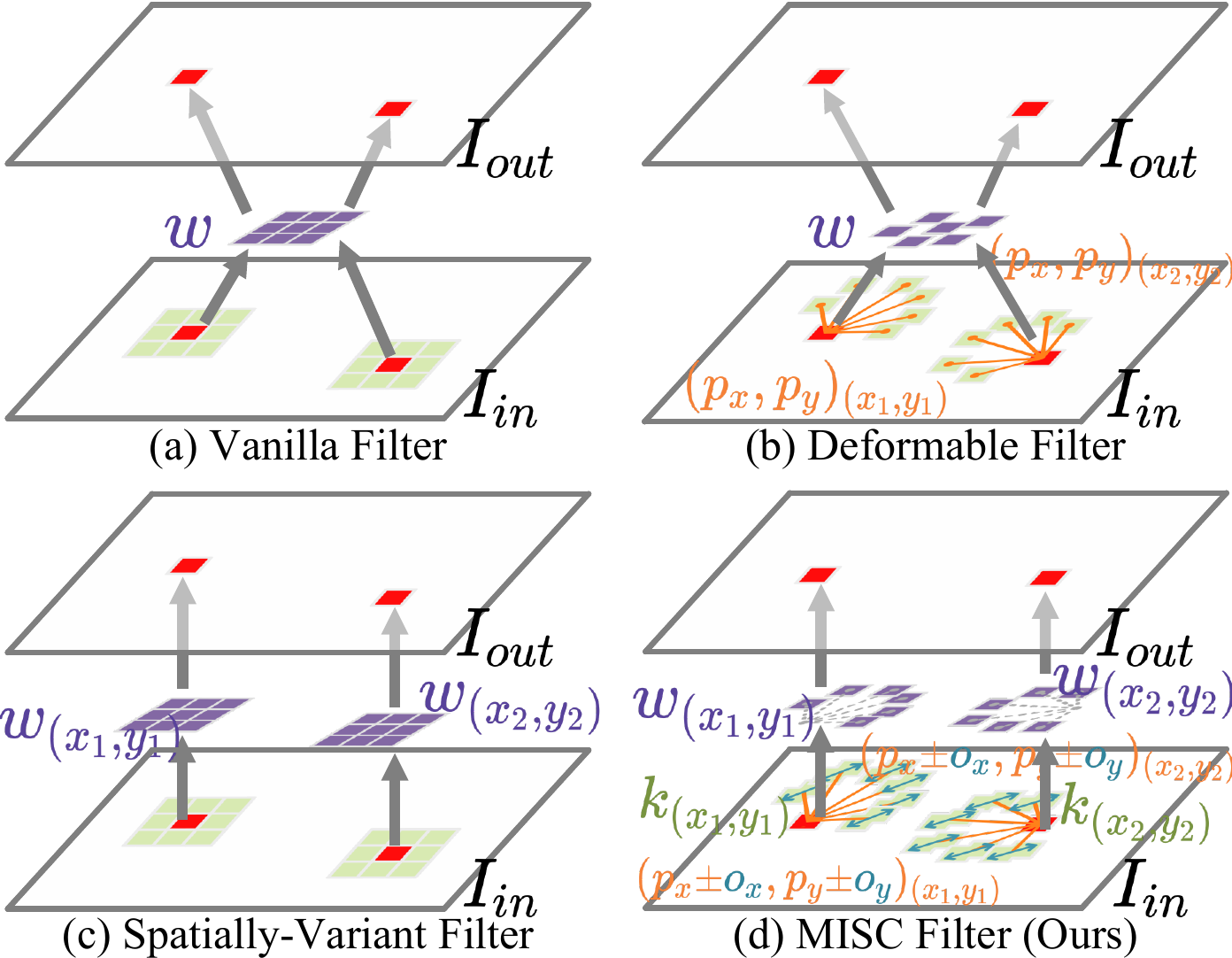}
    \vspace{-2mm}
   \caption{Illustration of other classical filters and our method. \textcolor{red}{Red} part is the center point of the filter in the image, and the \textcolor[RGB]{121,149,64}{green} part is the reference point for generating it. \textcolor{violet}{Violet} and \textcolor{orange}{orange} parts are the weights and offsets in the filter, respectively, {\color[RGB]{49,133,156} Blue} part of our method indicates the motion-guided alignment along the motion direction (best viewed in color).}
   \vspace{-2mm}
   \label{fig:teaser}
\end{figure}

\section{Introduction}
\label{sec:intro}

Blind motion deblurring aims at recovering high-quality images from the counterparts of blurred images containing real-world motions. 
From the perspective of the imaging process, real-world motions that produce blur are diverse and spatially varying, which introduces significant challenges for the corresponding solutions.

To tackle these issues, recent years have witnessed a growing number of image deblurring approaches, which can be categorized into two paradigms. 
The former attempts to strengthen the model capacity and directly reconstruct the residuals used to obtain sharp image~\cite{kong2023efficient,zamir2021multi,cho2021rethinking,tu2022maxim,wang2022uformer,zamir2022restormer,tsai2022stripformer,chen2022simple,mao2023intriguing,kupyn2018deblurgan,kupyn2019deblurgan}. 
These methods use multi-scale learning and supervision~\cite{zamir2021multi,cho2021rethinking,chen2022simple}, high efficiency attention mechanism~\cite{tu2022maxim,wang2022uformer,zamir2022restormer,tsai2022stripformer}, generative adversarial learning~\cite{kupyn2018deblurgan,kupyn2019deblurgan}, and frequency domain learning~\cite{kong2023efficient,mao2023intriguing,liu2023fsi} to optimize the model capability. 
Unfortunately, due to the shift-invariant of popularly used ``black box'' convolution networks, nearly all of these methods either lack interpretability to remove motion blur or the ability to handle spatially variable and large-scale motion. 
The latter attempts to learn the motion blur kernel or prior first, which then guides the reconstruction of residuals~\cite{sun2015learning,chen2023hierarchical,bahat2017non,chakrabarti2016neural,schuler2015learning,fang2023self}. 
These methods focus on designing different types of mathematical models to estimate motion prior, such as probabilistic distribution of motion blur~\cite{sun2015learning}, blur-field~\cite{bahat2017non}, latent representation of motion~\cite{chen2023hierarchical}, and motion kernel~\cite{schuler2015learning,chakrabarti2016neural,fang2023self}. 
These methods focus on reconstructing the image residuals and ignore the tailored handling of motion in image space, although promising results have been achieved.
Moreover, multiple factors in the real world (\eg,~camera shake, object large-scale movement, etc.) limit the accuracy of their predicted motion priors. Therefore, the image quality will be further improved if the complex motion in blurred images can be directly handled in the image space.

Different from reconstructing the residuals of sharp images in feature space via deep networks, reconstructing a high-quality image in image space usually involves various filtering operators~\cite{awate2006unsupervised}.
From easy to hard, vanilla filters (in Fig.~\ref{fig:teaser}(a)) are usually hand-crafted with a fixed shape operator for the whole image, limiting the ability to capture motion at far distances and at critical pixel features. Deformable filters~\cite{dai2017deformable} (in Fig.~\ref{fig:teaser}(b)) and spatially-variant filters~\cite{su2019pixel} (in Fig.~\ref{fig:teaser}(c)) further address these problems by adding spatial offsets and weights, respectively. We attempt to integrate the advantages of these filters to provide the ability for capturing various complex motions (in Fig.~\ref{fig:teaser}(d)). In addition, various filtering operators also have made significant progress in low-level vision recently, such as video artifact removal~\cite{lu2018deep}, video frame interpolation~\cite{lee2020adacof,cheng2021multiple}, multiple degradation removal~\cite{park2023all}, and so on. In summary, these demonstrate the generalizability and great potential of filtering for low-level restoration tasks.

To address the shortcomings of focusing only on reconstructing image residuals, we study the removal of motion blur in image space. 
Specifically, inspired by the deformable separable convolution~\cite{zhu2019deformable,dai2017deformable,wang2019edvr}, we propose a novel Motion-adaptive Separable Collaborative (MISC) Filter for blind motion deblurring. 
As shown in Fig.~\ref{fig:overview}, MISC Filter first aligns motion-induced blurring patterns to the middle along the predicted flow direction, and then collaboratively filters the alignment image to render the output. 
All parameters contained therein (\ie,~the motion field, mask, kernels, weights, and offsets) are predicted by a motion estimation network without affecting the reconstruction of residuals by existing methods. To improve the overall model efficiency, we further analyze the relationship between the motion estimation network and residual reconstruction network in an attempt to couple them.

The main contributions of this work are:
\begin{itemize}
    \item We propose a novel real-world deblurring filtering model called the Motion-adaptive Separable Collaborative (MISC) Filter. It targets the shortcomings of existing methods that focus only on image residual reconstruction in feature space and can handle more generalized and complex motion in image space.
    \item We analyze the relationship between the motion estimation network for producing filter parameters and the residual reconstruction network to maximize model efficiency.
    \item We demonstrate the effectiveness of our method by extensive quantitative and qualitative evaluations, and provide an effective solution for blind motion deblurring.
\end{itemize}

\section{Related Work}

\subsection{Blind Image Deblurring}
Blind motion deblurring is always a popular topic in low-level vision. Recently, to handle the degradation caused by complex motion patterns, an increasing number of methods have been proposed with significant success~\cite{chen2023hierarchical,wang2022uformer,tu2022maxim,zamir2022restormer,tsai2022stripformer,chen2022simple,mao2023intriguing,fang2023self,kong2023efficient}. According to whether the motion prior is required or not, these methods can be categorized into two groups, motion prior-free and motion prior-related. 

\vspace{-3mm}
\paragraph{Motion prior-free methods.} These methods~\cite{kong2023efficient,zamir2021multi,cho2021rethinking,tu2022maxim,wang2022uformer,zamir2022restormer,tsai2022stripformer,chen2022simple,mao2023intriguing,kupyn2018deblurgan,kupyn2019deblurgan,zhang2019deep,cui2023dual,cui2023focal,chen2021hinet,park2020multi,zhang2020deblurring,li2023efficient} focus attention on designing more robust feature learning networks that learn directly to remove various motion-induced blurring. In particular, these methods improve the model capacity from the perspectives of multi-scale learning and supervision~\cite{zamir2021multi,cho2021rethinking,chen2022simple,zhang2019deep,chen2021hinet,park2020multi}, high efficiency attention mechanism~\cite{tu2022maxim,wang2022uformer,zamir2022restormer,tsai2022stripformer,cui2023dual,cui2023focal,li2023efficient}, generative adversarial learning~\cite{kupyn2018deblurgan,kupyn2019deblurgan,zhang2020deblurring}, and frequency domain learning~\cite{kong2023efficient,mao2023intriguing}, separately. 
Typical MPRNet~\cite{zamir2021multi} and MIMO-Unet~\cite{cho2021rethinking} explore multi-scale and multi-stage constraint mechanisms, providing a robust framework for deblurring.
The latest FFTformer~\cite{kong2023efficient} optimizes the matrix multiplication in the Transformer by element-wise product in the frequency domain, significantly increasing the model capacity. Although significant progress has been made, the shift-invariant of the convolution limits the ability of these methods to deal with complex motion in the real world.

\begin{figure*}[t]
  \centering
   \includegraphics[width=0.94\linewidth,page=2]{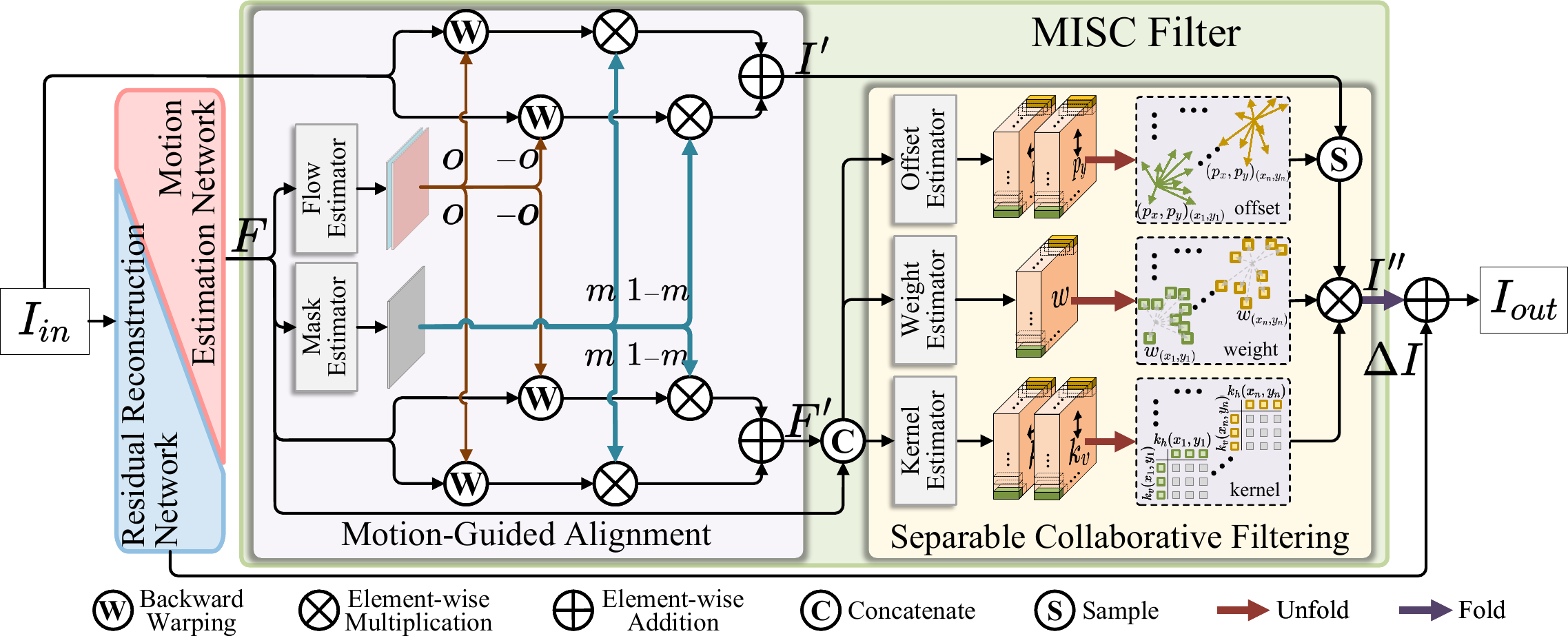}
   \vspace{-2mm}
   \caption{Overview of proposed Motion-adaptive Separable Collaborative (MISC) Filter. MISC Filter focuses on removing various motions in image space. It inputs image feature $F$ obtained from a motion estimation network and generates filtered image $I''$ via a motion-guided alignment (MGA) module and a separable collaborative filtering (SCF) module.}
   \label{fig:overview}
\end{figure*}

\vspace{-3mm}
\paragraph{Motion prior-related methods.} 
These methods~\cite{danielyan2011bm3d,purohit2021spatially,chi2021test,sun2015learning,chen2023hierarchical,bahat2017non,chakrabarti2016neural,schuler2015learning,fang2023self,liu2024decoupling,fei2023generative,xia2023diffir} treat them as inverse problems for motion patterns.
In particular, these methods first learn a spatially variable motion prior from the blurred appearance. The motion prior is then introduced into the reconstruction network to guide the network learning for motion blur removal. Typically, UFPNet~\cite{fang2023self} predicts the representations of motion blur through the Flow-based models and introduces them into residual reconstruction networks.
Inspired by the recent progress of Diffusion model~\cite{ho2020denoising}, Hi-Diff~\cite{chen2023hierarchical}, GDP~\cite{fei2023generative}, and DiffIR~\cite{xia2023diffir} estimate the motion prior and introduce them into the denoising process.
However, these methods emphasize the reconstruction of image residuals in feature space and ignore the direct handling of motions in image space. In addition, by adaptively estimating all parameters in the filter, our method is able to handle various complex motions.

\subsection{Filters in Low-level Vision}

Filtering algorithms have achieved widespread effect in various low-level tasks in recent years~\cite{danielyan2011bm3d,lu2018deep,lee2020adacof,cheng2021multiple,liu20234d,park2023all}. 
Typically, Lu~\etal~\cite{lu2018deep} introduces the Kalman filtering process into video artifact removal and builds a recursive filtering scheme. 
In the video frame interpolation task, AdaCoF~\cite{lee2020adacof} and EDSC~\cite{cheng2021multiple} propose an adaptive collaboration of flows and enhanced deformable separable convolution for intermediate frame synthesis by filtering operations, respectively.  
More recently, to deal with unknown multiple degradations in unpredictable realistic environments, ADMS~\cite{park2023all} proposes an adaptive discriminative filter-based model and achieves promising results. In general, due to the strong ability to deal with spatially-variant degradations, filters demonstrate great potential in low-level tasks. Therefore, in this paper, we introduce a MISC Filter to improve the ability to remove motion blur in the image space.

\section{Our Approach}

As shown in Fig.~\ref{fig:overview}, given a blurred input, our method uses a motion estimation network to obtain the parameters in the MISC Filter. The output sharp image is obtained from the combination of the filtered image by the MISC Filter and the image residuals from the reconstruction network. 

In this section, we first briefly introduce the problem formulation of the blind motion deblurring in Sec.~\ref{Problem Formulation} and then elaborate on the proposed MISC Filter in Sec.~\ref{MISC Filter}. Finally, we analyze the relationship between the motion estimation and the residual reconstruction network in Sec.~\ref{ncs}.

\subsection{Problem Formulation}
\label{Problem Formulation}
Blind motion deblurring aims to recover a sharp image from a blurred image without knowing the blur patterns. 
In this work, we define the blind motion deblurring task as an inverse problem for the blur kernel. 
Same as existing works~\cite{rim2020real,fang2023self}, we formulate a blurred image $\hat{y}$ as:
\begin{equation}
    I_{in} =k*I_{GT}+\tilde{n},
\end{equation}
where $I_{GT}$ is the sharp image. $k$ is the blur kernel produced by motion, which usually differs in each region depending on the variety of motions. $*$ denotes the filtering operations. $\tilde{n}$ denotes the additive noise of the camera. Based on this assumption, solving 1) the inverse problem of finding $k$ from a blurred image $I_{in}$ and 2) filtering the image directly in the image space are essential to recovering a sharp image.

\subsection{MISC Filter}
\label{MISC Filter}
Existing filters~\cite{dai2017deformable,cheng2020video,cheng2021multiple,su2019pixel} either fail to capture various spatially-variant motions or are limited by the degrees of freedom of multiple parameter choices (\eg,~kernels, weights, offsets, etc.). 
Therefore, as illustrated in Fig.~\ref{fig:overview}, we propose a Motion-adaptive Separable Collaborative (MISC) Filter to deal with these problems. 
Specifically, it first takes the image features learned from the motion estimation network as input. A motion-guided alignment (MGA) module then aligns the motion-induced blur to the motion middle along the estimated flow direction. A separable collaborative filtering (SCF) module finally uses the predicted filter parameters to filter the aligned image as output.

In terms of formula, given a blurred image $I_{in}$, we use a motion estimation network consisting of CNNs to obtain a feature denoted as $F\in \mathbb{R}^{C\times H\times W}$. $H$, $W$, and $C$ represent the feature's height, width, and channel, respectively. 
The $\mathrm{MGA}(\cdot)$ and $\mathrm{SCF}(\cdot)$ denote the MGA module and SCF module, respectively. The aligned image $I'\in \mathbb{R}^{3\times H\times W}$ and feature $F'\in \mathbb{R}^{C\times H\times W}$ are formulated as:
\begin{equation}
    I',F' = \mathrm{MGA}(I_{in},F).
\end{equation}
Then, the filtered image $I''\in \mathbb{R}^{3\times H\times W}$ of our MISC Filter can be formulated as:
\begin{equation}
    I'' = \mathrm{SCF}(I',F').
\end{equation}
The final output sharp image $I_{out}$ is obtained by summing the filtered image $I''$ with the image residual $\Delta I\in \mathbb{R}^{3\times H\times W}$ obtained from the reconstruction network. Following, we detail two core modules $\mathrm{MGA}(\cdot)$ and $\mathrm{SCF}(\cdot)$.

\vspace{-3mm}
\paragraph{Motion-guided alignment.}
Motion blur often occurs by the object displacement in a short period. To generate a sharper image, we propose the MGA module to localize the object's final position. 
First, a flow estimator is used to predict the motion field from motion start and end points in blurred images to the middle in latent sharp images. Then, bi-directional warping~\cite{liu2022learning,chan2021basicvsr} is utilized to align the blur to the middle moment, extending the range of handling blur and sharpening texture details.
In addition, to avoid the pixel occlusion~\cite{bao2019depth,liu2023ttvfi} in different directions during warping, we incorporate a estimator to generate mask as a modulation mechanism to optimize bi-directional pixel synthesis. 

Specifically, for the input feature $F$, we first obtain the motion flow and mask using the flow estimator $E_{f}(\cdot)$ and the mask estimator $E_{m}(\cdot)$, respectively, formulated as:
\begin{equation}
    \begin{aligned}
        o &= E_{f}(F),\\
        m &= E_{m}(F),
    \end{aligned}
\end{equation}
where $o\in \mathbb{R}^{2\times H\times W}$ is the motion flow, indicating the offset of each pixel along the direction from the motion center to the motion end. It is obtained from a flow estimator consisting of only one convolution layer. $m\in \mathbb{R}^{1\times H\times W}$ denotes the mask to adaptively adjust the weighted summation of pixels that are aligned to the motion center from the motion start and end. Different from the flow estimator, the output of the mask estimator is fed to a sigmoid function. Such a design not only enables the automatic parameter update of the estimator during training but also lighter the motion estimation in existing works~\cite{sun2018pwc,ilg2017flownet,dosovitskiy2015flownet}.

We then use bi-directional warping to obtain the aligned image $I'$, and the aligned feature $F'$, formulated as:
\begin{equation}
    \begin{aligned}
        &I' = m \otimes  \mathrm{W}(o,I_{in}) \oplus  (1-m) \otimes  \mathrm{W}(-o,I_{in}),\\
        &F' = m \otimes  \mathrm{W}(o,F) \oplus  (1-m) \otimes \mathrm{W}(-o,F),
    \end{aligned}
\end{equation}
where $\mathrm{W}(o,\cdot)$ denotes the backward warping~\cite{jaderberg2015spatial} according to the motion flow $o$. $\otimes$ and $\oplus$ denote the element-wise multiplication and element-wise addition, respectively. Following this, the obtained aligned feature are fed to the $\mathrm{SCF}(\cdot)$ module to further estimate the parameters that are used to filter the aligned image.

\vspace{-3mm}
\paragraph{Separable collaborative filtering.}

To alleviate the limitations of multiple degrees of freedom in filter parameter settings and find reference pixels when capturing motions, we collaboratively obtain the filter parameters by using kernel, weight, and offset estimators. The obtained parameters serve on the aligned image $I'$ to output a sharper result $I''$. 

Specifically, for the input feature $F$ and aligned feature $F'$, we separately use three estimators to predict the kernel, offset, and weight of the filter, formulated as:
\begin{equation}
    \begin{aligned}
        k_v,&k_h = E_{k}(C(F,F')),\\
        p_x,&p_y = E_{o}(C(F,F')),\\
        w &= E_{w}(C(F,F')).
    \end{aligned}
\end{equation}
Among them, inspired by the existing works~\cite{rigamonti2013learning,niklaus2017video,cheng2020video}, we approximate a 2D kernel with a pair of 1D kernels. This design encodes an $n\times n$ kernel with only $2n$ variables. $k_v\in \mathbb{R}^{n\times H\times W}$ and $k_h\in \mathbb{R}^{n\times H\times W}$ are the corresponding 1D filter kernels in the vertical and horizontal directions, respectively. $n$ denotes the kernel size. 
$w\in \mathbb{R}^{n^2\times H\times W}$ is the weight used to differentially aggregate different referenced pixels.
$p_x\in \mathbb{R}^{n^2\times H\times W}$ and $p_y\in \mathbb{R}^{n^2\times H\times W}$ are the offset of referenced pixels on the $x$-axis and $y$-axis, respectively. $C(\cdot)$ is the concatenation operation. $E_{k}(\cdot)$, $E_{o}(\cdot)$, and $E_{w}(\cdot)$ are lightweight estimators consisting of only one convolution layer with the same structure. The weight estimator additionally incorporates a softmax function to ensure that the pixel value after filtering is normalized to a value between 0 and 1. All parameters in the estimators are automatically learned during training.

\begin{figure*}[t]
  \centering
    \includegraphics[width=1.0\linewidth,page=3]{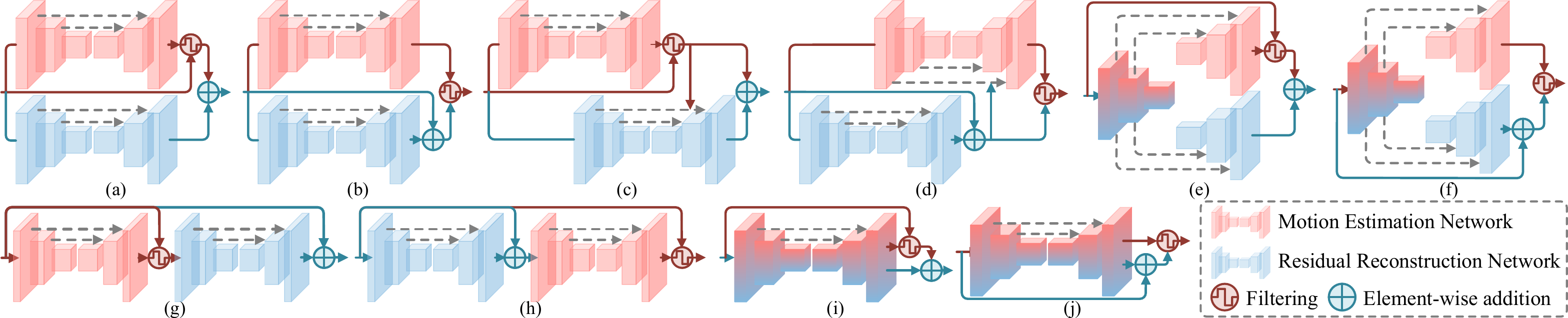}
    \vspace{-4mm}
   \caption{Design of different coupling strategies for motion estimation and residual reconstruction networks. In the ablation studies of Tab.~\ref{tab:f}, we explore different strategies to generate filters and residuals.}
   \label{fig:network}
\end{figure*}

Taking the pixel with coordinates $(a,b)$ in the filtered image $I''$ as example, the filtering process is formulated as:
\begin{equation}
    \begin{aligned}
    I''(a,b)= &\sum_{{\scriptsize\big (p_x(a,b),p_y(a,b)\big )}  }\bigg (w(a,b)\otimes k(a,b) \\
    &\otimes S\Big (I',\big (a{\small+}p_x(a,b),b{\small +}p_y(a,b)\big )\Big )\bigg ),
    \end{aligned}
\end{equation}
where $S\Big (I',\big (a{\small+}p_x(a,b),b{\small +}p_y(a,b)\big )\Big )$ represents the spatial sampling operation from the pixel in $I'$ with coordinates $\big (a{\small+}p_x(a,b),b{\small +}p_y(a,b)\big )$. $I''$ is the output filtered image. $k(a,b) = k_v(b)\cdot k_h^{T}(a)$ denotes the 2D kernel which can be approximated with two 1D kernels, effectively reducing the computational cost to a linear magnitude.

In summary, ``Kernel'' is the initial filter weight, which is generated adaptively depending on the blurred content of different regions. ``Offset'' is the vector direction of the motion that induces the local blur, and it locates the blurred boundary around the reconstructed pixel and captures the blur's shape. ``Weight'' implies a quadratic adjustment of content aggregation, increasing the model's non-linear representation. This design with collaborative parameter estimation significantly improves the ability to capture complex motions while avoiding cumbersome parameter settings.

\subsection{Network Coupling Strategies}
\label{ncs}
To estimate the filter parameters, we introduce a motion estimation network similar to the existing encoder-decoder structure~\cite{cho2021rethinking,mao2023intriguing} in our model. 
However, with a fixed model size, a large motion estimation network will improve the filtering performance but degrade the performance of the residual reconstruction, and vice versa. 
In addition, since filtering and residual reconstruction are completely independent, it is necessary to analyze the order of their execution.
Therefore, to trade-off between the motion estimation network and the residual reconstruction network, we analyze the relationship between them with different coupling strategies in an attempt to maximize the model efficacy with minimal model size.

Specifically, we analyze them in detail regarding 1) the coupling strategies of the two networks and 2) the order of filtering and reconstruction. As shown in Fig.~\ref{fig:network}(a)-(j), we build a total of ten frameworks.
Among them, we set up the following five coupling strategies:
\begin{itemize}
\item Parallel-based (Figs.~\ref{fig:network}(a) and (b)), where the feature learning of the two networks is completely independent. 
\item Semi-parallel-based (Figs.~\ref{fig:network}(c) and (d)), where the output of one is fed into the other as latent encoding.
\item Serial-based (Figs.~\ref{fig:network}(g) and (h)), where the output of one network serves as an input to the other one.
\item Semi-shared-based (Figs.~\ref{fig:network}(e) and (f)), where the encoders in both networks are shared.
\item Shared-based (Figs.~\ref{fig:network}(i) and (j)), where both networks share the same encoder-decoder structure and parameters.
\end{itemize}
For each strategy, the order of filtering and residual reconstruction can be exchanged.

For fair comparisons, we keep them with the same size to study the performance. Among them, the highest performance is achieved when based on the shared-based strategy and filtering first (in Fig.~\ref{fig:network}(i)). This is due to the mutual facilitation of the two parts of features used for filter parameters estimation and residual reconstruction. Specific experimental results can be obtained from Sec.~\ref{Framework.}.

\begin{table*}[t]\small
\centering
\begin{tabular}{lcccccccccc}
\toprule
\multirow{2}{*}{Method}     &  \multirow{2}{*}{\#P(M)}  &  \multirow{2}{*}{RT(s)}  & \multicolumn{2}{c}{RealBlur-R} & \multicolumn{2}{c}{RealBlur-J} & \multicolumn{2}{c}{GoPro} & \multicolumn{2}{c}{HIDE}    \\
\cmidrule[0.1pt](lr{0.125em}){4-5}\cmidrule[0.1pt](lr{0.125em}){6-7}\cmidrule[0.1pt](lr{0.125em}){8-9}\cmidrule[0.1pt](lr{0.125em}){10-11}
\myrowcolour%
 &  &  & PSNR & SSIM & PSNR & SSIM    & PSNR & SSIM & PSNR & SSIM  \\
\midrule
SRN~\cite{tao2018scale}           & 6.8     & 0.07      & 38.65&0.965     & 31.38&0.909   & 30.26 & 0.934     & 28.36&0.904      \\
\myrowcolour%
DeblurGANv2~\cite{kupyn2019deblurgan}   & 60.9     & 0.04           & 36.44&0.934     & 29.69&0.870   & 29.55& 0.934     & 26.61& 0.875        \\
SAPHN~\cite{suin2020spatially}      & 23     & 0.77           & -& -      & -& -    & 31.85& 0.948     & 29.98& 0.930      \\
\myrowcolour%
MPRNet~\cite{zamir2021multi}      & 20.1     & 0.09           & 39.31&0.972     & 31.76&0.922  & 32.66& 0.958     & 30.96& 0.939       \\
MIMO-Unet+~\cite{cho2021rethinking}      & 16.1     & 0.02           &  - & -     & 31.92&0.919   & 32.45& 0.956     & 29.99& 0.930     \\
\myrowcolour%
MAXIM~\cite{tu2022maxim}      & -     & -            & 39.45&0.962     & 32.84&0.935   & 32.86& 0.961     & 32.83& 0.956      \\
Uformer-B~\cite{wang2022uformer}      & 50.9     & 0.07           & -& -      & -& -   & 33.06& 0.967     & 30.90& 0.953     \\
\myrowcolour%
Restormer~\cite{zamir2022restormer}      & 26.1     & 0.08             & -& -      & -& -  & 32.92& 0.961     & 31.22& 0.942     \\
Restormer+local~\cite{chu2022improving}      & 26.1     & 0.42          & -& -      & -& -    & 33.57& 0.965     & 31.49& 0.944       \\
\myrowcolour%
MSDI-Net~\cite{li2022learning}      & 135.4     & -             & -& -      & -& -  & 33.28& 0.964     & 31.02 & 0.940      \\
Stripformer~\cite{tsai2022stripformer}      & 26.1     & 0.04            & 39.84&0.973     & 32.48&0.929   & 33.08& 0.962     & 31.03& 0.939      \\
\myrowcolour%
NAFNet~\cite{chen2022simple}      & 67.9     & 0.04          & -& -      & -& -    & 33.71& 0.966     & 31.31& 0.942      \\
DeepRFT+~\cite{mao2023intriguing}      & 23     & 0.09         & 39.84&0.972     & 32.19&0.930       & 33.23& 0.963     & 31.42 & 0.944    \\
\myrowcolour%
UFPNet~\cite{fang2023self}      & 80.3     & -         &  \textcolor{blue}{\underline{40.61}}&\textcolor{blue}{\underline{0.974}}     & \textcolor{blue}{\underline{33.35}}&\textcolor{blue}{\underline{0.934}}     & 34.06& \textcolor{blue}{\underline{0.968}}     & \textcolor{red}{31.74} & \textcolor{red}{0.947}       \\
FFTformer~\cite{kong2023efficient}      & 16.6     & 0.13          & 40.11&0.973     & 32.62&0.932    & \textcolor{red}{34.21}& \textcolor{red}{0.969}     & 31.62& 0.945      \\
\midrule
\myrowcolour%
\textbf{MISC Filter (Ours)}      & 16.0     & 0.07       & \textcolor{red}{41.23}     & \textcolor{red}{0.978}    & \textcolor{red}{33.88}     & \textcolor{red}{0.938}     & \textcolor{blue}{\underline{34.10}}     & \textcolor{red}{0.969}      & \textcolor{blue}{\underline{31.66}}      & \textcolor{blue}{\underline{0.946}} \\
\bottomrule
\end{tabular}
\vspace{-2mm}
\caption{Quantitative comparison on the RealBlur-R~\cite{rim2020real}, RealBlur-J~\cite{rim2020real}, GoPro~\cite{nah2017deep}, and HIDE~\cite{shen2019human} dataset. We follow existing works~\cite{kong2023efficient,mao2023intriguing} using a model trained on the RealBlur-R and RealBlur-J training sets for testing on the RealBlur-R and RealBlur-J test sets, respectively, and using a model trained on the GoPro training set for testing on the GoPro and HIDE test sets. \#P(M) and RT(s) indicate the parameter and runtime, respectively. Runtime is computed on images with the size of $256\times256$ with an Nvidia RTX 3090 GPU. \textcolor{red}{Red} and \textcolor{blue}{\underline{blue}} indicate the best and the second best performance, respectively (best viewed in color).}
\label{tab:tab1}
\end{table*}

\begin{figure*}[t]
  \centering
    \includegraphics[width=1.0\linewidth,page=4]{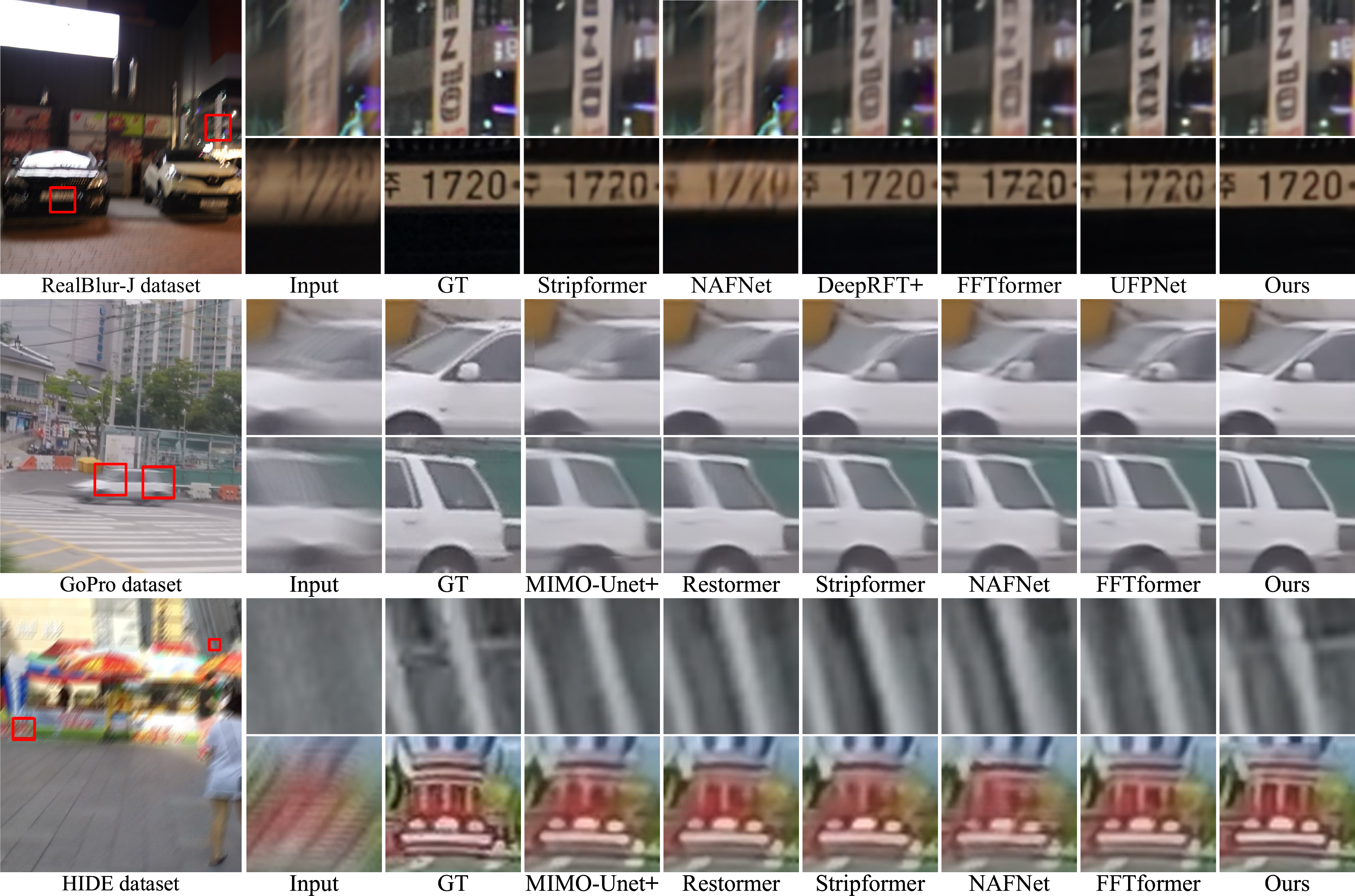}
    \vspace{-6mm}
   \caption{Visual results on \textbf{Realblur-J}~\cite{rim2020real}, \textbf{GoPro}~\cite{nah2017deep}, and \textbf{HIDE}~\cite{shen2019human} datasets. The method is shown at the bottom of each case. Zoom in to see better visualization.}
   \label{fig:case}
\end{figure*}

\section{Experiments}

\subsection{Datasets and Metrics}
We evaluate the our method on widely-used datasets: \textbf{RealBlur}~\cite{rim2020real}, \textbf{GoPro}~\cite{nah2017deep}, and \textbf{HIDE}~\cite{shen2019human}. \textbf{RealBlur}~\cite{rim2020real} dataset contains two subsets: \textbf{RealBlur-R} and \textbf{RealBlur-J}. Each subset contains 3,758 image pairs for training and 980 pairs for testing. 
\textbf{GoPro}~\cite{nah2017deep} dataset includes 2,103 pairs for training and 1,111 pairs for testing. \textbf{HIDE}~\cite{shen2019human} dataset only includes 2,025 images pairs for testing. 
For fair comparisons, we follow previous works~\cite{chen2023hierarchical,kong2023efficient,mao2023intriguing} to 1) train the model with the RealBlur-R and RealBlur-J training sets and test the model with their test sets, and 2) train the model with the GoPro training set and test the model with the test set of GoPro and HIDE. Same as previous works~\cite{kong2023efficient,zamir2021multi}, we keep the same evaluation metrics:1) Peak Signal-to-Noise Ratio (PSNR) and 2) Structural Similarity Index (SSIM)~\cite{wang2004image}.

\subsection{Implementation Details}

Similar to existing works~\cite{cho2021rethinking,zamir2021multi,mao2023intriguing}, the proposed filters are learned with supervision at multiple scales. 
To strengthen the capacity of the motion estimation network and the residual reconstruction network, we use the same U-Net structure, frequency-domain learning scheme, and loss function as prior works~\cite{kong2023efficient,mao2023intriguing}.
In SCF, the kernel size $n$ is set to 7. During training, we use Cosine Annealing scheme~\cite{loshchilov2016sgdr} and Adam~\cite{kingma2014adam} optimizer with $\beta_{1}=0.9$ and $\beta_{2}=0.999$. The learning rate is reduced from the initial $2\times 10^{-4}$ to $1\times 10^{-6}$. We set the batch size as $16$ and the input patch size as $256\times 256$ and augment the data with random horizontal flips, vertical flips, and $90^{\circ}$ rotations. The total number of epochs is 6K. All experiments were based on PyTorch and trained on 2 Nvidia V100 GPUs.

\subsection{Comparisons with State-of-the-art Methods}
We compare our method with 15 classical start-of-the-art methods. We summarize these methods into two categories: Motion prior-free~\cite{tao2018scale,kupyn2019deblurgan,zamir2021multi,cho2021rethinking,tu2022maxim,wang2022uformer,zamir2022restormer,tsai2022stripformer,chen2022simple,mao2023intriguing,kong2023efficient,chu2022improving,li2022learning} and Motion prior-related~\cite{tsai2022banet,fang2023self}. For fair comparisons, we report the results from their original papers or reproduce the results through the published models.

\vspace{-2mm}
\paragraph{Quantitative comparison.}
The performance comparison of our method with other SOTA methods is shown in Tab.~\ref{tab:tab1}. Compared to the latest motion prior-free method FFTformer~\cite{kong2023efficient}, our method achieves significant performance gains on the more complex real-world motion blur datasets RealBlur-R~\cite{rim2020real} and RealBlur-J~\cite{ rim2020real}. While our method achieves comparable performance on GoPro~\cite{nah2017deep} and HIDE~\cite{shen2019human}, it only spends half the runtime. It proves the robustness of our method to handle complex real-world motions. Compared to the latest motion prior-related method UFPNet~\cite{fang2023self}, our method achieves higher performance on the GoPro~\cite{nah2017deep}, RealBlur-R~\cite{rim2020real}, and RealBlur-J~\cite{rim2020real} with less number of parameters and runtime. It proves the superiority of our method in handling various motions and verifies the generalization ability on various motion blurs.

\vspace{-2mm}
\paragraph{Qualitative comparison.}
To further compare the visual quality of different methods, we present their de-blurring results in Fig.~\ref{fig:case}.
For fair comparisons, we directly use author-released models to get results for fairness. The results of our method have better visual quality, especially when complex motion is involved. For example, in the first case of Fig.~\ref{fig:case}, our method recovers the clear text with complex movement. In the second case, our method recovers the texture details of a fast-moving car. 
More visual results can refer to the supplementary materials.

\subsection{Ablation study}
In this section, we conduct the ablation on the proposed MISC Filter and analyze the network coupling strategies. In addition, we further analyze the effect of the different filtering methods and kernel sizes. All experiments are performed on our lightweight version model on GoPro~\cite{nah2017deep}.

\begin{figure}[t]
  \centering
    \includegraphics[width=1.0\linewidth,page=5]{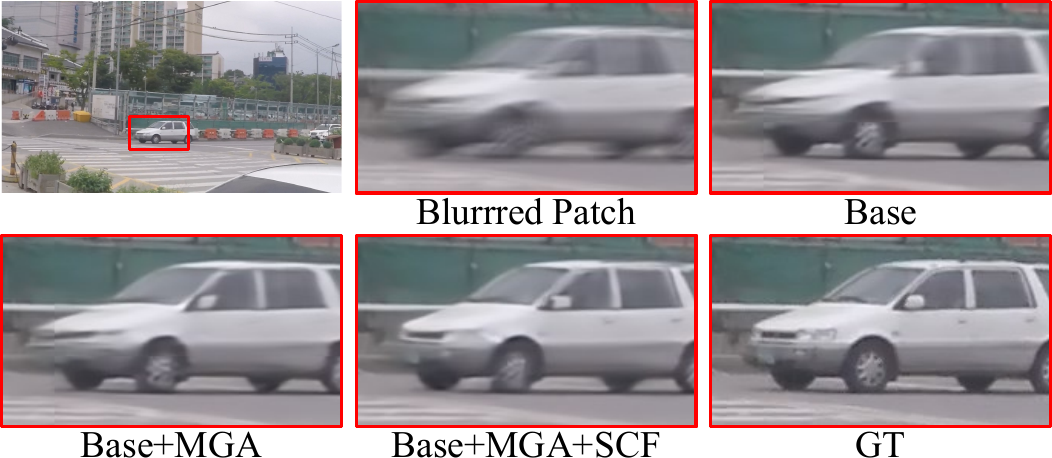}
    \vspace{-5mm}
   \caption{Visualization of ablation studies on MISC Filter.}
   \label{fig:abl}
\end{figure}

\vspace{-2mm}
\paragraph{MISC Filter.}
To demonstrate the superiority of each component in our MISC Filter, we analyze their effect on performance in Tab.~\ref{tab:abl} and Fig.~\ref{fig:abl}. We try to keep the same model size in each comparison for fairness. We directly remove the MISC Filter as the ``Base'' model and progressively add each component for comparisons. 
The addition of MGA improves PSNR (from 32.40 dB to 32.51 dB) and visual quality, verifying its ability to eliminate motion blur through bi-directional alignment. Adding the kernel, weight, and offset estimators separately in the SCF can all lead to a 0.1 dB improvement, demonstrating the superiority of each component in our MISC Filter. With all of them added, the model performance improves from 32.51dB to 32.83dB, and the visual quality is further improved. It proves that our method alleviates the limitations of multiple degrees of freedom in filter parameters setting and enhances the ability to handle various complex motions.

\begin{table}[t]\small
\centering
\begin{tabular}{ccccccc}
\toprule
\multirow{2}{*}{Base}  & \multirow{2}{*}{MGA}  & \multicolumn{3}{c}{SCF}   & \multirow{2}{*}{PSNR}  & \multirow{2}{*}{SSIM}   \\
\cmidrule[0.1pt](lr{0.125em}){3-5}
\myrowcolour%
  &   &  kernel & weight &  offset&     &   \\
\midrule
\checkmark  &  & & &           &  32.40  &  0.955 \\  
\myrowcolour%
\checkmark  & \checkmark& & &   & 32.51  &   0.956 \\   
\checkmark  & \checkmark&  \checkmark& &   & 32.55  & 0.957   \\
\myrowcolour%
\checkmark  & \checkmark&  \checkmark& \checkmark&   & 32.67  &  0.957  \\
\checkmark  & \checkmark&  \checkmark& &  \checkmark & 32.71  &   0.958 \\
\myrowcolour%
\checkmark &  &  \checkmark&  \checkmark&  \checkmark & 32.67  &   0.958 \\
\checkmark  &\checkmark&\checkmark&\checkmark&\checkmark& \textbf{32.83}  &  \textbf{0.960}   \\
\bottomrule
\end{tabular}
\vspace{-2mm}
\caption{Results of ablation studies on MISC Filter. MGA: motion-guided alignment. SCF: separable collaborative filtering. our MISC Filter can be interpreted as ``Base+MGA+SCF''.}
\vspace{-2mm}
\label{tab:abl}
\end{table}

\begin{figure}[t]
  \centering
    \includegraphics[width=1.0\linewidth,page=6]{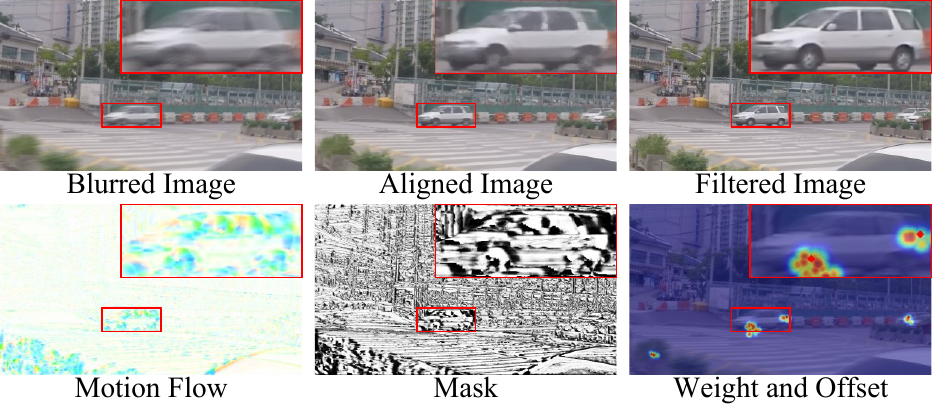}
    \vspace{-5mm}
   \caption{Visualization of the aligned image following MGA, motion flow, mask, weight, and offset in MISC Filter.}
   \label{fig:vis}
\end{figure}

To demonstrate the model ability of the filtering process to capture complex motions, we further visualize the aligned image following MGA, motion flow, mask, weight, and offset in the MISC Filter. As shown in Fig.~\ref{fig:vis}, the motion flow and mask can effectively capture the moving car, and the aligned image following MGA removes the blurring induced by fast motion. 
In addition, we chose pixels of the wheels and back of the car (indicated by red dots) to visualize the weights and offsets, where stronger signals indicate higher weights. It demonstrates the strong ability of the filter to capture non-local spatial motions.

\vspace{-3mm}
\paragraph{Network coupling strategies.}
\label{Framework.}
To analyze the relationship between the motion estimation network and the residual reconstruction network in Sec.~\ref{ncs}, we compare all ten frameworks presented in Fig.~\ref{fig:network}. In Tab.~\ref{tab:f}, the highest performance is achieved when the two networks share the same structure and when filtering is performed before adding the residuals (Group (i) in Fig.~\ref{fig:network}). This is because sharing the same network would facilitate interactive learning of the two-part features. Moreover, prioritizing filtering blurred images in image space is more straightforward for resolving motion blur. Therefore, we use the framework (i) to maximize model efficacy with minimal model size.

\begin{table}[t]\small
\centering
\begin{tabular}{c|>{\columncolor[gray]{0.925}} cc >{\columncolor[gray]{0.925}} cc  >{\columncolor[gray]{0.925}}c}
\toprule
Group & a&  b& c & d&  e\\  
\midrule
PSNR &  32.53 &  32.42 & 32.56& 32.38 & 32.63\\
SSIM &  0.958 & 0.956 &  0.958 & 0.956 & 0.959\\
\midrule
Group & f &  g&  h & i & j \\
\midrule
PSNR &  32.59 & 32.55 & 32.46& \textbf{32.83} & 32.39 \\
SSIM & 0.958 &  0.957 &  0.957 & \textbf{0.960} & 0.956\\
\bottomrule
\end{tabular}
\vspace{-2mm}
\caption{Results of ablation studies on frameworks in Fig.~\ref{fig:network}.}
\label{tab:f}
\end{table}

\vspace{-2mm}
\paragraph{Filtering.}
To demonstrate the reliability of our filtering process, we compare several classical filtering methods~\cite{dai2017deformable,su2019pixel,cheng2021multiple} in Tab.~\ref{tab:fil}. Our MISC Filter performs favorably against the two representative filters, deformable filter~\cite{dai2017deformable} and separable filter~\cite{su2019pixel}. It is because they either cannot adaptively change the weights of different regions or are not conducive to capturing non-local motion. In addition, due to the design of its motion-guided alignment module and collaborative kernel estimation, our method has higher performance than deformable separable filter~\cite{cheng2021multiple}. Our method allows for handling more complex motions.

\begin{table}[t]\small
\centering
\begin{tabular}{lcc}
\toprule
Method    & PSNR  & SSIM   \\
\midrule
\myrowcolour%
Vanilla Filter &  32.21  & 0.952 \\
Deformable Filter~\cite{dai2017deformable} &   32.48  &  0.956 \\ 
\myrowcolour%
Separable Filter~\cite{su2019pixel} &   32.51  &  0.957 \\
Deformable Separable Filter~\cite{cheng2021multiple} &   32.54  &  0.957 \\
\midrule
\myrowcolour%
\textbf{MISC Filter (Ours)} &   \textbf{32.83} &  \textbf{0.960} \\
\bottomrule
\end{tabular}
\vspace{-2mm}
\caption{Results of ablation studies on different filtering methods.}
\label{tab:fil}
\end{table}

\vspace{-2mm}
\paragraph{Kernel size.}
We analyze the effect of kernel size $n$ on describing complex motions for the MISC Filter. 
As shown in Tab.~\ref{tab:ks}, the performance positively correlates with kernel size. It demonstrates the powerful potential of our MISC Filter to handle complex motions. However, the performance gain gradually decreases when the $n$ exceeds 7. Besides, although a larger kernel will allow more pixels to be referenced, it will also increase the number of parameters.

\begin{table}[t]\small
\centering
\begin{tabular}{c| >{\columncolor[gray]{0.925}} cc >{\columncolor[gray]{0.925}} cc >{\columncolor[gray]{0.925}} c}
\toprule
$n$ & 3 &  5 & 7 & 9&  11\\
\midrule
PSNR & 32.60 &32.76 &32.83 & 32.84& 32.87\\
SSIM &  0.957 & 0.958 & 0.960 &0.960 & 0.961\\
\bottomrule
\end{tabular}
\vspace{-2mm}
\caption{Results of ablation studies on kernel size $n$.}
\label{tab:ks}
\end{table}

\subsection{Evaluation on hardware-induced blurring}
Aside from motion-induced blurring, in the under-display camera (UDC) systems, the pixel array of light-emitting diodes used for display diffracts and attenuates the incident light, resulting in blurring~\cite{koh2022bnudc}. Therefore, we construct experiments on widely-used UDC benchmark T-OLED~\cite{zhou2021image} in order to explore the ability of our method to remove hardware-induced blurring. As shown in Tab.~\ref{tab:oled}, our method achieves comparable performance compared to most models specifically designed for this task. However, our method still falls short since it cannot handle the low-light problem caused by the attenuation of incident light in this task.

\begin{table}[t]\small
\centering
\begin{tabular}{lcccc}
\toprule
Method & RT(s) & \#P(M) & PSNR & SSIM  \\
\midrule
\myrowcolour%
MSUNet~\cite{zhou2021image}           & 0.08     & 8.9        & 37.40                & 0.9756        \\
DAGF~\cite{sundar2020deep}             & 1.12     & 1.1          & 36.49                & 0.9716        \\
\myrowcolour%
PDCRN~\cite{panikkasseril2020transform}            & 0.08     & 4.7    & 37.83                & 0.9780         \\
ResUNet~\cite{yang2021residual}         & 0.43      & 16.5    & 37.95                & 0.9790           \\
\myrowcolour%
RDUNet~\cite{yang2021residual}           & 0.53     & 47.9   & 38.13                & 0.9797          \\
UDC-UNet~\cite{liu2022udc}           & 0.30     & 5.7    & 38.10                & 0.9796      \\
\myrowcolour%
BNUDC~\cite{koh2022bnudc}           & 0.08      & 4.6   & 38.22                & 0.9798    \\
FSI~\cite{liu2023fsi}           & 0.07      & 3.8   & \textcolor{red}{38.60}                & \textcolor{red}{0.9805}        \\
\midrule
\myrowcolour%
\textbf{MISC Filter (Ours)}             &  0.05 & 5.6 &  \textcolor{blue}{\underline{38.42}}  & \textcolor{blue}{\underline{0.9800}} \\
\bottomrule
\end{tabular}
\vspace{-2mm}
\caption{Results of hardware-induced blur removal.}
\label{tab:oled}
\end{table}

\section{Conclusion}
In this paper, we introduce a new perspective to handle motion blur in image space instead of features and propose a novel motion-adaptive separable collaborative (MISC) filter. 
In the MISC Filter, we design a motion-guided alignment module and a separable collaborative filtering module to eliminate motion-induced blur. To achieve this, we introduce an additional motion estimation network to predict the filter parameters and further analyze the coupling strategies between it and the conventional residual reconstruction network. 
Such a design eliminates blurring induced by complex motions while avoiding cumbersome filter parameter settings.
Experimental results show its advantages and the potential to remove blur from other scenarios.

\paragraph{Acknowledgement.} This work was supported in part by the NSFC under Grant 62272380 and 62103317, the Fundamental Research Funds for the Central Universities, China (xzy022023051), the Innovative Leading Talents Scholarship of Xi'an Jiaotong University, and MEGVII.

\clearpage
{
    \small
    \bibliographystyle{ieeenat_fullname}
    \bibliography{main,ref}
}

\end{document}